\documentclass[%
twocolumn,
amsmath,amssymb,
aps, physrev,
]{revtex4-2}
\usepackage[utf8]{inputenc}
\usepackage[T1]{fontenc}
\usepackage{graphicx}
\usepackage{dcolumn}
\usepackage{bm}
\usepackage{xcolor}

\begin{document}

\preprint{APS/123-QED}

\title{Reconciling GW170817 and GW190814 with a Nonmonotonic Sound-Speed Equation of State}
\author{Marlon M. S. Mendes}
\email{marlon.mendes.102043@ga.ita.br}
\affiliation{Departamento de F\'isica e Laborat\'orio de Computa\c{c}\~ao Cient\'ifica Avan\c{c}ada e Modelamento (Lab-CCAM), Instituto Tecnol\'ogico de Aeron\'autica, DCTA, 12228-900, S\~ao Jos\'e dos Campos, SP, Brazil}

\author{Odilon Louren\c{c}o da Silva Filho}
\affiliation{Departamento de F\'isica e Laborat\'orio de Computa\c{c}\~ao Cient\'ifica Avan\c{c}ada e Modelamento (Lab-CCAM), Instituto Tecnol\'ogico de Aeron\'autica, DCTA, 12228-900, S\~ao Jos\'e dos Campos, SP, Brazil}

\author{C\'esar H. Lenzi}
\affiliation{Departamento de F\'isica e Laborat\'orio de Computa\c{c}\~ao Cient\'ifica Avan\c{c}ada e Modelamento (Lab-CCAM), Instituto Tecnol\'ogico de Aeron\'autica, DCTA, 12228-900, S\~ao Jos\'e dos Campos, SP, Brazil}
\date{\today}

\begin{abstract}
We show that the GW170817--GW190814 tension can be reconciled within General Relativity by a structured causal EoS basin found with a Constrained Evolutionary TOV Discovery pipeline (CETD). The basin contains $\sim1.4\times10^4$ unique EoS with $M_{\max}=2.3$--$2.8\,M_\odot$, $R_{1.4}=11.97$--$12.29$ km, $\Lambda_{1.4}^{\rm exact}\le580$, and double-peaked sound speeds reaching $c_s^2/c^2=0.86$--$0.99$.
\end{abstract}

\maketitle

\section{Introduction}

The multimessenger events GW170817 and GW190814 impose apparently conflicting requirements on the neutron star equation of state (EoS): GW170817 favors comparatively soft matter in the tidal regime, while GW190814 may require an unusually stiff high-density response if its secondary is interpreted as a neutron star \cite{Abbott2017GW170817,Abbott2018GW170817EOS,Capano2020,Abbott2020GW190814}. GW230529 further sharpens this context by reinforcing the relevance of compact objects in the lower mass gap \cite{Abac2024GW230529}. The central question is therefore no longer whether this tension is recognized \cite{Tan2020,Fattoyev2020,Nathanail2021,Annala2020,Kojo2021}, but whether a physically admissible region of EoS space can accommodate these observations simultaneously within standard General Relativity.

The difficulty is well known. An EoS stiff enough to support 
$\sim 2.6\,M_\odot$ comes into tension with independent constraints 
from heavy-ion collisions and from the tidal deformabilities of 
medium-mass stars~\cite{Fattoyev2020,YaoEtAl2024StructureSpeedSoundHIC}, whereas one soft enough to 
satisfy GW170817 typically limits the maximum mass to 
$M_{\rm TOV}\lesssim 2.3\,M_\odot$~\cite{Nathanail2021}. 
NICER measurements~\cite{Gendreau2016,Miller2021,Riley2021} and 
low-density nuclear constraints~\cite{Reed2021,Fattoyev2018} further restrict the allowed 
pressure profile. If a viable solution exists, it must therefore be strongly density 
dependent: soft in the tidal regime, yet rapidly stiffening in the 
deep core.

This expectation is consistent with recent analyzes showing that the sound speed of cold dense matter need not vary monotonically, but may contain peaks, troughs, plateaus, or multi-peak structures at supranuclear densities \cite{Zhou2024TwoPeaksSoundSpeedQCD,CaiLi2024StrongGravitySoundSpeedPeaks,MroczekEtAl2024NontrivialFeaturesSoundSpeed,EckerRezzolla2022ScaleIndependentSoundSpeed}. In particular, Bayesian evidence for a two-peak sound-speed scenario becomes especially relevant if the $\sim2.6\,M_\odot$ component of GW190814 is interpreted as a neutron star \cite{Zhou2024TwoPeaksSoundSpeedQCD}.

Rather than proposing a specific microphysical mechanism, we ask whether the allowed region has an identifiable geometry in EoS space-searching not for a single exceptional model, but for a robust basin of viable solutions. To this end, we introduce a \emph{Constrained Evolutionary TOV Discovery} pipeline (CETD), a staged evolutionary search combining many-objective candidate generation, Tolman-Oppenheimer-Volkoff (TOV) screening, causal and observational filtering, and elitist refinement around validated seeds. CETD is used to determine whether viable branches exist outside the Read piecewise-polytropic region \cite{Read2009} and whether they persist under progressively stricter physical and observational filters. The novelty of the present work is therefore not the statement that these events are in tension, but the demonstration that their combined observational pressure maps onto a structured and recurrent basin of EoS solutions within standard General Relativity.

The paper is organized as follows. We first summarize the observational constraints, then describe the CETD exploration and basin analysis, and finally examine the physical content of the recovered branches through their mass-radius, pressure-energy-density, and sound-speed behavior under TOV selection.
\section{Search Strategy}

We explore neutron-star EoS space beyond the standard
piecewise-polytropic sector of Read~\textit{et al.}~\cite{Read2009},
allowing each adiabatic index to vary over $\Gamma_i\in(0,10)$.
The search is performed with the Constrained Evolutionary TOV
Discovery pipeline (CETD), which couples staged evolutionary
generation/refinement to numerical TOV integration
\cite{Tolman1939,OppenheimerVolkoff1939}. At each stage, candidates
are screened by causality, compactness, tidal deformability, radius,
maximum-mass, and observational filters; validated survivors seed the
next refinement step.

The CETD selection vector contains eight objectives: (i) causality
$v_s^2 \equiv c_s^2/c^2 \leq 1$; (ii) maximum-mass support
$M_{\max}\geq2\,M_\odot$; (iii) canonical-radius agreement; (iv)
tidal-deformability agreement with GW170817; parameter-space novelty
in (v) $\Gamma_i$ and (vi) $\log_{10}P_1$; and observable-space
novelty in (vii) $M$--$R$ and (viii) $\Lambda$.

The search proceeds in three stages. Stage~1 tests \emph{existence}:
a broad beyond-Read exploration identifies whether any physically
admissible EoS survives outside the Read interval, followed by a
local basin expansion around the first viable seed. Stage~2 tests
\emph{mass amplification}: Stage~1 survivors are locally refined to
increase $M_{\max}$ while preserving compact $R_{1.4}$ and the
GW170817 tidal bound $\Lambda_{1.4}^{\rm exact}<580$
\cite{Abbott2018GW170817EOS,Abbott2019GW170817}. Stage~3 tests
\emph{observational realization}: a deeper TOV-in-the-loop refinement
is compared against NICER constraints
\cite{Miller2019,Riley2019,Miller2021,Riley2021}, massive-pulsar
lower bounds on $M_{\max}$
\cite{Demorest2010,Antoniadis2013,Cromartie2020,Fonseca2021}, and
the high-mass scale motivated by GW190814~\cite{Abbott2020GW190814}.

After the direct TOV, causal, mass, radius, and tidal filters, we
apply three \emph{a posteriori} diagnostics. Bootstrap resampling tests
whether the recovered basin is stable rather than a sampling
fluctuation~\cite{Efron1979,EfronTibshirani1993}. An I--Love relation test
checks consistency with universal relations between moment of inertia
and tidal deformability
\cite{Hinderer2008,YagiYunes2013Science,YagiYunes2013PRD}. Finally,
the Komoltsev--Kurkela diagnostic~\cite{KomoltsevKurkela2022} tests
whether each recovered EoS admits a causal and thermodynamically
stable connection to a high-density pQCD anchor. Thus, the validation
targets seven hypotheses: non-empty beyond-Read existence, support of
the target $M_{\max}$ window, strict causality, consistency with the
GW170817 tidal bound, bootstrap stability, I--Love consistency, and
\emph{a posteriori} pQCD connectability.
\section{Results}

We tested whether the GW170817--GW190814 tension can be addressed
within General Relativity by an extended EoS basin rather than a
single fine-tuned solution. The CETD search proceeded in three
stages: existence under the beyond-Read constraint
($\Gamma_i\in(0,10)$, beyond the Read et al.~\cite{Read2009}
interval $\Gamma_i\in[2,4]$); mass amplification of the Stage~1
survivors while preserving compact $R_{1.4}$ and
$\Lambda_{1.4}^{\rm exact}<580$; and observational realization
across the recovered $M_{\max}\in(2.3,2.8)\,M_\odot$ window.

\begin{figure*}[t]
    \centering
    \includegraphics[width=0.98\textwidth]{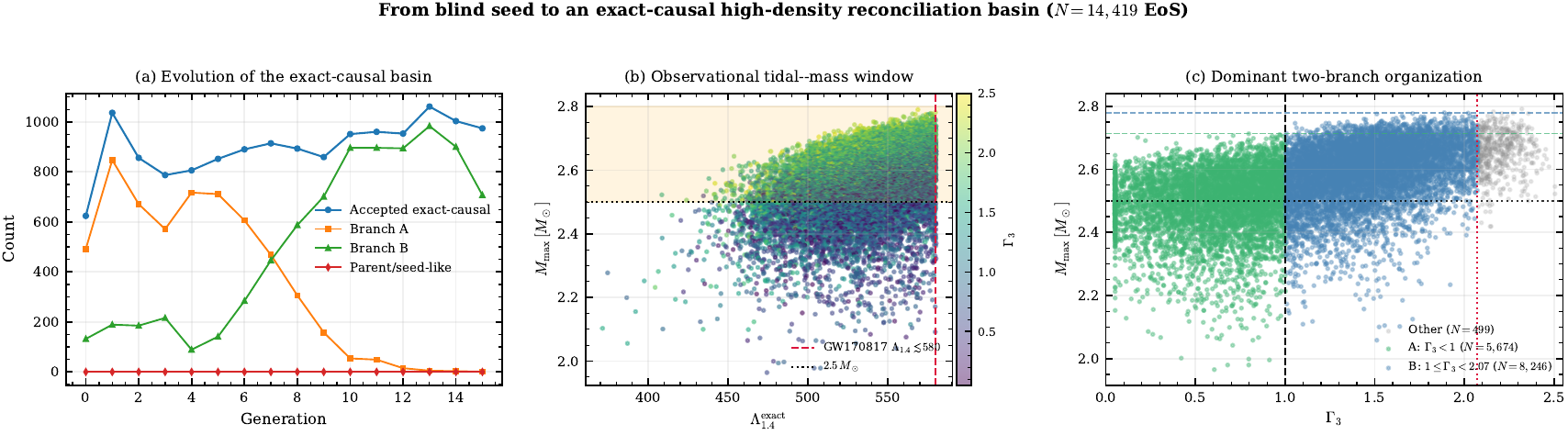}
    \caption{
    Stage~2 mass amplification.
    (a) Generational screening of the candidate pool through 
    TOV, causal, astrophysical, and parent-selection filters.
    (b) Causal survivors in the $\Lambda_{1.4}^{\rm exact}$--$M_{\max}$ 
    plane. The dashed line marks $\Lambda_{1.4}^{\rm exact}=580$ and 
    the shaded band marks the GW190814 mass window.
    (c) The same survivors in the $\Gamma_3$--$M_{\max}$ plane, 
    separating into two branches: Branch~A ($\Gamma_3<1$) and 
    Branch~B ($\Gamma_3\geq1$). Both branches reach high $M_{\max}$ 
    while remaining compatible with the low-tidal-deformability 
    constraint.
    }
    \label{fig:discovery}
\end{figure*}

The Stage~1 search isolates an initial viable EoS beyond the
Read et al.~\cite{Read2009} sector at the high-$\Gamma_1$ edge of
the explored domain, demonstrating that admissible solutions are
rare but non-empty. A local exploration around this seed reveals a
first basin, which CETD expands through TOV-in-the-loop refinement
into a multi-family population from which Stage~2 seeds are drawn.
Stage~2 shows that mass amplification does not collapse under the
TOV, causal, and astrophysical filters: the surviving solutions
populate the same $\Lambda_{1.4}$--$M_{\max}$ window and separate
into two branches in the $\Gamma_3$--$M_{\max}$ plane
(Fig.~\ref{fig:discovery}), a softening branch
(A: $\Gamma_3<1$, $N=5{,}674$, $M_{\max}\le 2.71\,M_\odot$) and a
stiff hadronic branch (B: $\Gamma_3\ge 1$, $N=7{,}643$,
$M_{\max}\le 2.79\,M_\odot$). Both reach the GW190814-compatible
mass range, indicating that the reconciliation is not a single
fine-tuned solution but an extended two-branch basin.

\begin{figure*}[t]
    \centering
    \includegraphics[width=0.70\textwidth]{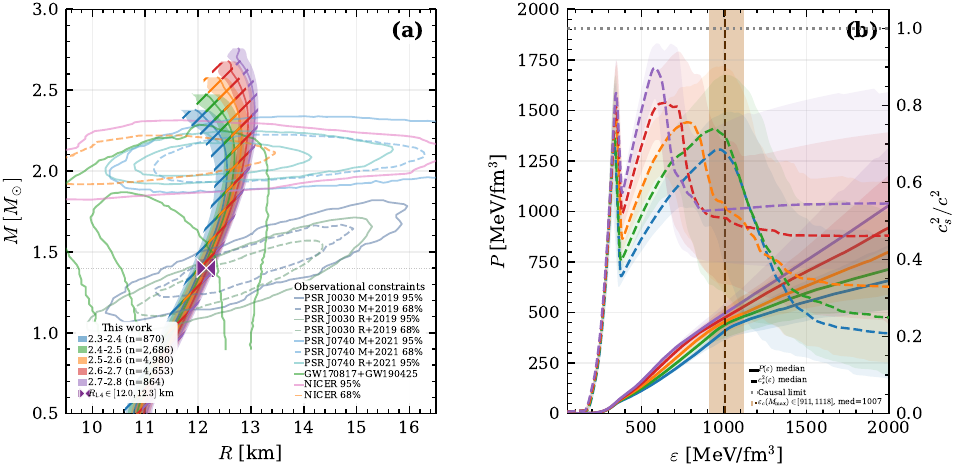}
\caption{
Causal equation-of-state solutions grouped by maximum-mass bin.
Panel~(a) shows the corresponding mass--radius sequences for five 
$M_{\max}$ intervals spanning $2.3$--$2.8\,M_\odot$, overlaid on 
multimessenger observational constraints from NICER 
(PSR~J0030+0451 and PSR~J0740+6620) and gravitational-wave 
measurements (GW170817+GW190425). The marked interval at 
$M=1.4\,M_\odot$ indicates the compact-radius band selected by 
the recovered solutions, $R_{1.4} \in (11.97, 12.29)$~km.
Panel~(b) shows the associated median pressure 
$P(\varepsilon)$ and squared sound speed $v_s^2(\varepsilon)/c^2$ 
for the same bins, with shaded bands indicating the corresponding 
$10$--$90\%$ dispersion. The vertical orange band marks the 
central energy-density interval at maximum mass, 
$\varepsilon_c(M_{\max}) \in (911, 1118)$~MeV/fm$^3$, with 
median $\varepsilon_c \simeq 1007$~MeV/fm$^3$. The horizontal 
dotted line denotes the causal limit $v_s^2=1$.
}
    \label{fig:physics}
\end{figure*}

Stage~3 projects the recovered basin onto the mass--radius plane
(Fig.~\ref{fig:physics}(a)), confirming compatibility with
multimessenger constraints across the displayed
$M_{\max}\in(2.3,2.8)\,M_\odot$ window while maintaining compact
canonical radii in $R_{1.4}\in(11.97,12.29)$~km. The corresponding
$P(\varepsilon)$ and $v_s^2(\varepsilon)$ profiles
(Fig.~\ref{fig:physics}(b)) reveal a robust double-peaked
sound-speed structure: a sharp first peak at intermediate densities,
followed by partial softening, and a broader second peak that
develops across the central energy-density band of the maximum-mass
configurations, $\varepsilon_c(M_{\max})\in(911,1118)\,{\rm MeV/fm^3}$.
Depending on the EoS, the central state samples either the rising
flank or the summit of this second peak, so the stiffening
responsible for high-mass support is precisely the response probed
in the densest regions of the star. This behavior is qualitatively
aligned with recent evidence that resolving the GW170817--GW190814
tension favors a strongly density-dependent and non-monotonic
sound-speed profile~\cite{Zhou2024TwoPeaksSoundSpeedQCD,
CaiLi2024StrongGravitySoundSpeedPeaks,
MroczekEtAl2024NontrivialFeaturesSoundSpeed,
EckerRezzolla2022ScaleIndependentSoundSpeed}. The solutions remain
causal ($v_s^2\le 1$) throughout, peaking well above the $1/3$
conformal bound~\cite{Bedaque2015}.

Table~\ref{tab:representative_eos} positions our result relative to
previous works that established that a neutron-star interpretation
of the $\sim 2.6\,M_\odot$ GW190814 secondary requires a highly
nontrivial high-density response, hard to reconcile with nucleonic
EoS under multimessenger constraints~\cite{Tan2020,Fattoyev2020,
Nathanail2021}. We confirm this requirement but change its
interpretation: rather than imposing a microscopic phase transition,
exotic degree of freedom, or predefined sound-speed morphology, the
required nontrivial response emerges dynamically from the CETD
search once the beyond-Read restriction on $\{\Gamma_i\}$ is relaxed.

\begin{table*}[t]
\caption{%
Representative causal EoS in each \(M_{\max}\) bin after
TOV-in-the-loop selection. The displayed binned population contains
\(N=14{,}053\) unique causal configurations in
\(M_{\max}\in(2.3,2.8)\,M_\odot\). All listed configurations satisfy
strict causality, \(c_s^2/c^2<1\), and the joint
GW170817+NICER+pulsar multimessenger constraints.%
}
\label{tab:representative_eos}
\centering
\scriptsize
\setlength{\tabcolsep}{4.0pt}
\renewcommand{\arraystretch}{1.12}

\begin{ruledtabular}
\begin{tabular}{lccccccccccc}
\(M_{\max}\) bin
& \(N\)
& \(M_{\max}\)
& \(R_{1.4}\)
& \(\Lambda_{1.4}^{\rm exact}\)
& \(v_s^2\)
& \(\log_{10}P_1\)
& \(\rho_1^\dagger\)
& \(\rho_2^\dagger\)
& \(\Gamma_1\)
& \(\Gamma_2\)
& \(\Gamma_3\)
\\
\multicolumn{1}{c}{\([M_\odot]\)}
& \multicolumn{1}{c}{}
& \multicolumn{1}{c}{\([M_\odot]\)}
& \multicolumn{1}{c}{\([{\rm km}]\)}
& \multicolumn{1}{c}{}
& \multicolumn{1}{c}{}
& \multicolumn{1}{c}{\([{\rm dyn\,cm^{-2}}]\)}
& \multicolumn{1}{c}{\(\dagger\)}
& \multicolumn{1}{c}{\(\dagger\)}
& \multicolumn{1}{c}{}
& \multicolumn{1}{c}{}
& \multicolumn{1}{c}{}
\\
\hline
\(2.3\!-\!2.4\) & 870   & 2.370 & 12.06 & 517 & 0.858 & 34.916 & 0.000948 & 0.002331 & 6.910 & 2.507 & 0.725 \\
\(2.4\!-\!2.5\) & 2686  & 2.464 & 12.07 & 519 & 0.910 & 34.951 & 0.000958 & 0.002461 & 6.919 & 2.628 & 0.717 \\
\(2.5\!-\!2.6\) & 4980  & 2.567 & 12.15 & 535 & 0.930 & 34.993 & 0.000968 & 0.002095 & 6.568 & 2.755 & 1.042 \\
\(2.6\!-\!2.7\) & 4653  & 2.636 & 12.20 & 547 & 0.963 & 35.048 & 0.000981 & 0.001488 & 6.467 & 3.021 & 1.518 \\
\(2.7\!-\!2.8\) & 864   & 2.791 & 12.27 & 569 & 0.988 & 35.050 & 0.000966 & 0.001453 & 6.623 & 3.303 & 2.314 \\
\end{tabular}
\end{ruledtabular}

\vspace{0.1em}
\begin{minipage}{0.96\textwidth}
\footnotesize
\(\dagger\) Transition densities \(\rho_1\) and \(\rho_2\) are reported
in the internal geometrized units of the TOV/EoS solver, with
\(G=c=1\). The pressure scale is reported as
\(\log_{10}(P_1/{\rm dyn\,cm^{-2}})\). Dimensionless columns are
\(\Lambda_{1.4}^{\rm exact}\), \(v_s^2\), and \(\Gamma_i\).
\end{minipage}

\end{table*}

The basin is therefore not a single exceptional point but a
structured sequence of high-quality survivors spanning the tested
$M_{\max}$ bins, populating a narrow band with
$\Lambda_{1.4}^{\rm exact}\in(371,580)$ and sound-speed profiles
that approach but do not exceed the relativistic bound. Across the
corrected set of $N\simeq 1.4\times10^4$ unique causal EoS, the
GW170817 tidal bound is preserved throughout, the inferred
$\Lambda_{1.4}$ range is consistent with GW170817~\cite{
Abbott2018GW170817EOS,Capano2020} and with multimessenger Bayesian
posteriors that incorporate heavy-ion, nuclear-theory, and kilonova
data~\cite{Huth2022MicroscopicMacroscopic,Pang2023NMMA}. The
recovered density-dependent response is therefore compatible with
methodologically independent inferences while providing a
constructive realization of the corresponding high-density mechanism.

\paragraph*{Robustness.}
The basin survives a comprehensive set of audits. The double- or
multi-peaked $v_s^2(\varepsilon)$ morphology is exhibited by
$95.61\%$ of the basin ($13{,}786/14{,}419$) and is stable under
changes in derivative resolution, transition-edge masking, and
peak-prominence threshold ($0.9247$--$0.9760$ across the robustness
grid), with $v_s^2\simeq 0.86$--$0.99$, well above the conformal
limit $1/3$~\cite{Bedaque2015} and below the exact causal bound.
This is consistent with the response identified by Tan
\textit{et al.}~\cite{Tan2020} as necessary for a $\sim 2.6\,M_\odot$
neutron star, but here it emerges dynamically once the 
Read-like restriction $\Gamma_i\in[2,4]$ is relaxed. An independent
RK4 TOV reintegration of $500$ representative EoS reproduces stored
observables within $|\Delta M_{\max}|\le 2.25\times 10^{-3}\,M_\odot$
and $|\Delta R_{1.4}|\le 4.70\times 10^{-2}$~km
(95th-percentiles), excluding integration artifacts. The basin is
also not a consequence of a finely tuned cut: simultaneous cuts
$\Lambda_{1.4}\le 580$, $M_{\max}\ge 2.5\,M_\odot$,
$R_{1.4}\in[11.5,13.5]$~km leave $72.8\%$ of the basin, and
progressively tighter causal margins $v_s^2\le 0.99,\,0.98,\,0.95$
still leave $12{,}893$, $11{,}610$, and $8{,}076$ EoS, with the
maximum-mass frontier reaching $M_{\max}=2.744\,M_\odot$ even at
$v_s^2\le 0.95$, excluding the interpretation that the result
relies on numerical pinning to the causal bound. The basin is
recurrent across $137$ source seeds and $16$ generations
(largest seed $6.75\%$, effective number $57.4$); leave-one-seed
and leave-one-generation tests preserve at least $92.45\%$ of the
population and $M_{\max}=2.779\,M_\odot$. The two branches are
separable in physical-parameter space (logistic-regression and
random-forest AUC~$\gtrsim 0.92$), with leading separators $\rho_2$,
$\Gamma_2$, $R(M_{\max})$, $M_{\max}$ and
$\Delta R(M_{\max})\simeq 0.35$--$0.36$~km, while remaining nearly
degenerate at $1.4\,M_\odot$ --- showing that the branching is
controlled by the high-density stiffening scale rather than by an
imposed label on $\Gamma_3$.

\paragraph*{Stability and a posteriori connectability.}
All selected configurations satisfy the turning-point condition
$dM/d\rho_c>0$ up to the maximum-mass configuration~\cite{Sorkin1981,
Friedman1988}, and explicit solution of the Sturm--Liouville radial
eigenvalue problem~\cite{Chandrasekhar1964,Bardeen1966,Kokkotas2001}
on a representative subset confirms $\omega_0^2>0$ at canonical,
intermediate, and near-maximum masses, with $\omega_0^2$ decreasing
toward the turning point as expected. The pQCD-connectability
criterion of Komoltsev and Kurkela~\cite{KomoltsevKurkela2022} is
satisfied across a representative sample of $500$ basin EoS, with
connecting-EoS slopes
$\Delta P/\Delta\mu\in(0.93,5.25)\,{\rm fm}^{-3}$ in the allowed
window $(n_c,n_{\rm pQCD})\simeq(0.77,6.4)\,{\rm fm}^{-3}$; we use
this as a conservative consistency check rather than a hard prior.
A low-density anchoring test, replacing the sector below
$1.3\,n_0$ with a microphysical SLy4/CompOSE crust~\cite{
Douchin2001SLy,Typel2015CompOSE}, preserves TOV integrability and
high-mass support, showing that $M_{\max}$ is controlled by the
high-density continuation; without joint reoptimization of the
high-density sector, however, the hybrid models become too stiff
in $\Lambda_{1.4}$ and local $v_s^2$, and a dedicated follow-up is
required to construct a fully anchored EoS family under joint
$M_{\max}$, $R_{1.4}$, $\Lambda_{1.4}$, causality, radial-stability,
and pQCD-matching constraints.
\section{Conclusion}

In contrast to modified-gravity interpretations of the GW190814
secondary~\cite{Moffat2020MOGGW190814,
Astashenok2020ExtendedGravityGW190814,
Nunes2020ScreeningGravityGW190814,
Lin2022FTGravityGW190814,
Charmousis2022EGBGW190814,
Reyes2024PostTOVScreenedGW190814}, we show that the
GW170817--GW190814 tension can be resolved within standard General
Relativity. The joint observations constrain the admissible
high-density EoS space, selecting a well-populated causal basin of
\(14{,}419\) equations of state that simultaneously accommodate
canonical tidal softness, NICER-scale radii, massive-pulsar support,
and \(M_{\max}\simeq 2.3\)--\(2.8\,M_\odot\), all within the unmodified
Tolman--Oppenheimer--Volkoff framework~\cite{Abbott2017GW170817,
Abbott2018GW170817EOS,Abbott2020GW190814,Tolman1939,
OppenheimerVolkoff1939}. The physical message is sharp: within the
tested effective EoS class, dense matter must remain soft at canonical
densities and reorganize its stiffness rapidly at higher densities.
This response cannot be reproduced by a uniformly stiff EoS~\cite{
Bedaque2015,Annala2020}; instead, it emerges in more than \(95\%\) of
the basin as a double- or multi-peaked \(v_s^2(\varepsilon)\)
profile~\cite{Tan2020,MroczekEtAl2024NontrivialFeaturesSoundSpeed,
Zhou2024TwoPeaksSoundSpeedQCD,CaiLi2024StrongGravitySoundSpeedPeaks}.
The basin therefore exposes not one exceptional model but a structural
requirement. The central lesson is not that one can tune an EoS to
pass the current filters, but that the filters themselves carve out a
recurrent high-density geometry in which canonical softness and
extreme-mass support coexist only through a nonmonotonic sound-speed
response.

The result survives the principal ways it could have failed.
Independent TOV reintegrations reproduce the stored observables, the
peak morphology is stable under changes in the detection prescription,
and leave-one-out and bootstrap audits confirm that no small subset
controls the geometry. The two-branch internal structure reflects a
physical separation rather than a labeling artifact: the branches are
separable by \(\rho_2\), \(\Gamma_2\), \(R(M_{\max})\), and
\(M_{\max}\), yet remain nearly degenerate at canonical mass. The
configurations satisfy the turning-point criterion along the stable
branch, and explicit radial-pulsation audits find \(\omega_0^2>0\) on
the tested canonical, intermediate, and near-maximum-mass
models~\cite{Sorkin1981,Friedman1988,Chandrasekhar1964,Bardeen1966,
Kokkotas2001}. The basin is therefore a causal, TOV-selected,
stability-audited region of EoS space rather than a numerical
collection of large masses.

The proposal is falsifiable. It fails if future observations robustly
exclude neutron-star maximum masses above \(\sim2.4\,M_\odot\), drive
\(\Lambda_{1.4}\) below the surviving range, or contradict the
predicted branch-dependent separation in \(R(M_{\max})\). It is
strongly supported if a neutron star is found that combines extreme
gravitational mass with GW170817-compatible canonical tidal softness.
The present basin is a high-density reconciliation mechanism, not its
microscopic completion. The pQCD-connectability test is satisfied
across the audited sample, and the low-density anchoring test preserves
the basin's mass--radius structure under hybridization with a
microphysical crust, showing that the mechanism survives physically
motivated extensions~\cite{KomoltsevKurkela2022,Douchin2001SLy,
Typel2015CompOSE}. A joint reoptimization of the high-density sector
under \(M_{\max}\), \(R_{1.4}\), \(\Lambda_{1.4}\), causality, radial
stability, and pQCD matching is the natural next step toward a fully
anchored EoS family.

Independently of that microscopic anchoring, the double-peaked,
two-branch, stability-audited structure identified here is a concrete
prediction for next-generation gravitational-wave observatories and
post-merger spectroscopy~\cite{EinsteinTelescope2020,
CosmicExplorer2019,Bauswein2019,Bernuzzi2020}. These instruments will
either support the mechanism as a viable high-density route through
the GW170817--GW190814 tension or replace it with a more complete
theory of dense matter.
\section*{Acknowledgments}
The authors acknowledge support from the Instituto Tecnológico de Aeronáutica (ITA) and thank the Coordenação de Aperfeiçoamento de Pessoal de Nível Superior (CAPES), Brazil, for financial support. It is also supported by the Conselho Nacional de Desenvolvimento Cient\'ifico e Tecnol\'ogico (CNPq) under Grants No. 305327/2023-2 (C.H.L.), No. 401565/2023-8~(Universal - C.H.L.), No. 409736/2025-2~(Universal - C.H.L.), and Funda\c{c}\~ao de Amparo \`a Pesquisa do Estado de S\~ao Paulo (FAPESP) under Thematic Project No. 2024/17816-8 (C.H.L.).

\bibliography{apssamp}

@article{Abbott2017GW170817,
  author       = {Abbott, B. P. and others},
  title        = {{GW170817}: Observation of Gravitational Waves from a Binary Neutron Star Inspiral},
  journal      = {Physical Review Letters},
  volume       = {119},
  pages        = {161101},
  year         = {2017},
  doi          = {10.1103/PhysRevLett.119.161101}
}

@article{Abbott2018GW170817EOS,
  author       = {Abbott, B. P. and others},
  title        = {{GW170817}: Measurements of Neutron Star Radii and Equation of State},
  journal      = {Physical Review Letters},
  volume       = {121},
  pages        = {161101},
  year         = {2018},
  doi          = {10.1103/PhysRevLett.121.161101}
}

@article{Abbott2020GW190814,
  author       = {Abbott, R. and others},
  title        = {{GW190814}: Gravitational Waves from the Coalescence of a 23 $M_{\odot}$ Black Hole with a 2.6 $M_{\odot}$ Compact Object},
  journal      = {The Astrophysical Journal Letters},
  volume       = {896},
  number       = {2},
  pages        = {L44},
  year         = {2020},
  doi          = {10.3847/2041-8213/ab960f}
}

@article{Abac2024GW230529,
  author        = {Abac, A. G. and others},
  collaboration = {LIGO Scientific Collaboration and Virgo Collaboration and KAGRA Collaboration},
  title         = {Observation of Gravitational Waves from the Coalescence of a 2.5--4.5 {$M_\odot$} Compact Object and a Neutron Star},
  journal       = {The Astrophysical Journal Letters},
  volume        = {970},
  number        = {2},
  pages         = {L34},
  year          = {2024},
  doi           = {10.3847/2041-8213/ad5beb}
}

@article{Read2009,
  author        = {Read, Jocelyn S. and Lackey, Benjamin D. and Owen, Benjamin J. and Friedman, John L.},
  title         = {Constraints on a phenomenologically parameterized neutron-star equation of state},
  journal       = {Physical Review D},
  volume        = {79},
  pages         = {124032},
  year          = {2009},
  doi           = {10.1103/PhysRevD.79.124032},
  eprint        = {0812.2163},
  archivePrefix = {arXiv},
  primaryClass  = {astro-ph}
}

@article{Capano2020,
  author  = {Capano, Collin D. and Tews, Ingo and Brown, Stephanie M. and Margalit, Ben and De, Soumi and Kumar, Sumit and Brown, Duncan A. and Krishnan, Badri and Reddy, Sanjay},
  title   = {Stringent constraints on neutron-star radii from multimessenger observations and nuclear theory},
  journal = {Nature Astronomy},
  volume  = {4},
  pages   = {625--632},
  year    = {2020},
  doi     = {10.1038/s41550-020-1014-6}
}

@article{Tan2020,
  author  = {Tan, Hung and Noronha-Hostler, Jacquelyn and Yunes, Nico},
  title   = {Neutron Star Equation of State in Light of {GW190814}},
  journal = {Physical Review Letters},
  volume  = {125},
  pages   = {261104},
  year    = {2020},
  doi     = {10.1103/PhysRevLett.125.261104}
}

@article{Fattoyev2020,
  author  = {Fattoyev, F. J. and Horowitz, C. J. and Piekarewicz, J. and Reed, Brendan},
  title   = {{GW190814}: Impact of a 2.6 Solar Mass Neutron Star on the Nucleonic Equations of State},
  journal = {Physical Review C},
  volume  = {102},
  pages   = {065805},
  year    = {2020},
  doi     = {10.1103/PhysRevC.102.065805}
}

@article{Nathanail2021,
  author  = {Nathanail, Antonios and Most, Elias R. and Rezzolla, Luciano},
  title   = {{GW170817} and {GW190814}: Tension on the Maximum Mass},
  journal = {The Astrophysical Journal Letters},
  volume  = {908},
  number  = {2},
  pages   = {L28},
  year    = {2021},
  doi     = {10.3847/2041-8213/abdfc6}
}

@article{Miller2021,
  author  = {Miller, M. C. and Lamb, F. K. and Dittmann, A. J. and others},
  title   = {The Radius of {PSR J0740+6620} from {NICER} and {XMM-Newton} Data},
  journal = {The Astrophysical Journal Letters},
  volume  = {918},
  number  = {2},
  pages   = {L28},
  year    = {2021},
  doi     = {10.3847/2041-8213/ac089b}
}

@article{Riley2021,
  author  = {Riley, Thomas E. and Watts, Anna L. and Ray, Paul S. and others},
  title   = {A {NICER} View of the Massive Pulsar {PSR J0740+6620} Informed by Radio Timing and {XMM-Newton} Spectroscopy},
  journal = {The Astrophysical Journal Letters},
  volume  = {918},
  number  = {2},
  pages   = {L27},
  year    = {2021},
  doi     = {10.3847/2041-8213/ac0a81}
}

@article{Reed2021,
  author  = {Reed, Brendan T. and Fattoyev, Farrukh J. and Horowitz, Charles J. and Piekarewicz, Jorge},
  title   = {Implications of {PREX-2} on the Equation of State of Neutron-Rich Matter},
  journal = {Physical Review Letters},
  volume  = {126},
  pages   = {172503},
  year    = {2021},
  doi     = {10.1103/PhysRevLett.126.172503}
}

@article{Fattoyev2018,
  author  = {Fattoyev, F. J. and Piekarewicz, J. and Horowitz, C. J.},
  title   = {Neutron Skins and Neutron Stars in the Multimessenger Era},
  journal = {Physical Review Letters},
  volume  = {120},
  pages   = {172702},
  year    = {2018},
  doi     = {10.1103/PhysRevLett.120.172702}
}

@article{Bedaque2015,
  author  = {Bedaque, Paulo and Steiner, Andrew W.},
  title   = {Sound velocity bound and neutron stars},
  journal = {Physical Review Letters},
  volume  = {114},
  pages   = {031103},
  year    = {2015},
  doi     = {10.1103/PhysRevLett.114.031103}
}

@article{Annala2020,
  author  = {Annala, Eemeli and Gorda, Tyler and Kurkela, Aleksi and N{\"a}ttil{\"a}, Joonas and Vuorinen, Aleksi},
  title   = {Evidence for Quark-Matter Cores in Massive Neutron Stars},
  journal = {Nature Physics},
  volume  = {16},
  pages   = {907--910},
  year    = {2020},
  doi     = {10.1038/s41567-020-0914-9}
}

@article{Kojo2021,
  author  = {Kojo, Toru},
  title   = {QCD Equations of State and Speed of Sound in Neutron Stars},
  journal = {AAPPS Bulletin},
  volume  = {31},
  pages   = {11},
  year    = {2021},
  doi     = {10.1007/s43673-021-00011-6}
}

@article{KomoltsevKurkela2022,
  author        = {Komoltsev, Oleg and Kurkela, Aleksi},
  title         = {How Perturbative {QCD} Constrains the Equation of State at Neutron-Star Densities},
  journal       = {Physical Review Letters},
  volume        = {128},
  pages         = {202701},
  year          = {2022},
  doi           = {10.1103/PhysRevLett.128.202701},
  eprint        = {2111.05350},
  archivePrefix = {arXiv},
  primaryClass  = {nucl-th}
}

@article{Huth2022MicroscopicMacroscopic,
  author  = {Huth, Sabrina and Pang, Peter T. H. and Tews, Ingo and Dietrich, Tim and Le F{\`e}vre, Arnaud and Schwenk, Achim and Trautmann, Wolfgang and Agarwal, Kshitij and Bulla, Mattia and Coughlin, Michael W. and Van Den Broeck, Chris},
  title   = {Constraining Neutron-Star Matter with Microscopic and Macroscopic Collisions},
  journal = {Nature},
  volume  = {606},
  pages   = {276--280},
  year    = {2022},
  doi     = {10.1038/s41586-022-04750-w}
}

@article{Pang2023NMMA,
  author  = {Pang, Peter T. H. and Dietrich, Tim and Coughlin, Michael W. and others},
  title   = {An Updated Nuclear-Physics and Multi-Messenger Astrophysics Framework for Binary Neutron Star Mergers},
  journal = {Nature Communications},
  volume  = {14},
  pages   = {8352},
  year    = {2023},
  doi     = {10.1038/s41467-023-43932-6}
}

@inproceedings{Gendreau2016,
  author    = {Gendreau, Keith C. and Arzoumanian, Zaven and Adkins, Phillip W. and others},
  title     = {The {Neutron} Star {Interior} {Composition} {Explorer} ({NICER}): Design and Development},
  booktitle = {Space Telescopes and Instrumentation 2016: Ultraviolet to Gamma Ray},
  editor    = {den Herder, Jan-Willem A. and Takahashi, Tadayuki and Bautz, Marshall},
  series    = {Proc. SPIE},
  volume    = {9905},
  pages     = {99051H},
  year      = {2016},
  doi       = {10.1117/12.2231304}
}

@article{Efron1979,
  author  = {Efron, Bradley},
  title   = {Bootstrap Methods: Another Look at the Jackknife},
  journal = {The Annals of Statistics},
  volume  = {7},
  number  = {1},
  pages   = {1--26},
  year    = {1979},
  doi     = {10.1214/aos/1176344552}
}

@book{EfronTibshirani1993,
  author    = {Efron, Bradley and Tibshirani, Robert J.},
  title     = {An Introduction to the Bootstrap},
  publisher = {Chapman and Hall},
  address   = {New York},
  year      = {1993}
}

@article{Hinderer2008,
  author        = {Hinderer, Tanja},
  title         = {Tidal Love Numbers of Neutron Stars},
  journal       = {The Astrophysical Journal},
  volume        = {677},
  pages         = {1216--1220},
  year          = {2008},
  doi           = {10.1086/533487},
  eprint        = {0711.2420},
  archivePrefix = {arXiv},
  primaryClass  = {astro-ph}
}

@article{YagiYunes2013Science,
  author        = {Yagi, Kent and Yunes, Nicol{\'a}s},
  title         = {{I-Love-Q}: Unexpected Universal Relations for Neutron Stars and Quark Stars},
  journal       = {Science},
  volume        = {341},
  number        = {6144},
  pages         = {365--368},
  year          = {2013},
  doi           = {10.1126/science.1236462},
  eprint        = {1302.4499},
  archivePrefix = {arXiv},
  primaryClass  = {gr-qc}
}

@article{YagiYunes2013PRD,
  author        = {Yagi, Kent and Yunes, Nicol{\'a}s},
  title         = {{I-Love-Q} Relations in Neutron Stars and their Applications to Astrophysics, Gravitational Waves, and Fundamental Physics},
  journal       = {Physical Review D},
  volume        = {88},
  pages         = {023009},
  year          = {2013},
  doi           = {10.1103/PhysRevD.88.023009},
  eprint        = {1303.1528},
  archivePrefix = {arXiv},
  primaryClass  = {gr-qc}
}

@article{Tolman1939,
  author  = {Tolman, Richard C.},
  title   = {Static Solutions of Einstein's Field Equations for Spheres of Fluid},
  journal = {Physical Review},
  volume  = {55},
  pages   = {364--373},
  year    = {1939},
  doi     = {10.1103/PhysRev.55.364}
}

@article{OppenheimerVolkoff1939,
  author  = {Oppenheimer, J. R. and Volkoff, G. M.},
  title   = {On Massive Neutron Cores},
  journal = {Physical Review},
  volume  = {55},
  pages   = {374--381},
  year    = {1939},
  doi     = {10.1103/PhysRev.55.374}
}

@article{Miller2019,
  author        = {Miller, M. C. and Lamb, F. K. and Dittmann, A. J. and others},
  title         = {{PSR J0030+0451} Mass and Radius from {NICER} Data and Implications for the Properties of Neutron Star Matter},
  journal       = {The Astrophysical Journal Letters},
  volume        = {887},
  number        = {1},
  pages         = {L24},
  year          = {2019},
  doi           = {10.3847/2041-8213/ab50c5},
  eprint        = {1912.05705},
  archivePrefix = {arXiv},
  primaryClass  = {astro-ph.HE}
}

@article{Riley2019,
  author        = {Riley, T. E. and Watts, A. L. and Bogdanov, S. and others},
  title         = {A {NICER} View of {PSR J0030+0451}: Millisecond Pulsar Parameter Estimation},
  journal       = {The Astrophysical Journal Letters},
  volume        = {887},
  number        = {1},
  pages         = {L21},
  year          = {2019},
  doi           = {10.3847/2041-8213/ab481c},
  eprint        = {1912.05702},
  archivePrefix = {arXiv},
  primaryClass  = {astro-ph.HE}
}

@article{Demorest2010,
  author  = {Demorest, Paul B. and Pennucci, Tim and Ransom, Scott M. and Roberts, Mallory S. E. and Hessels, Jason W. T.},
  title   = {A Two-Solar-Mass Neutron Star Measured Using Shapiro Delay},
  journal = {Nature},
  volume  = {467},
  pages   = {1081--1083},
  year    = {2010},
  doi     = {10.1038/nature09466}
}

@article{Antoniadis2013,
  author        = {Antoniadis, John and Freire, Paulo C. C. and Wex, Norbert and others},
  title         = {A Massive Pulsar in a Compact Relativistic Binary},
  journal       = {Science},
  volume        = {340},
  number        = {6131},
  pages         = {1233232},
  year          = {2013},
  doi           = {10.1126/science.1233232},
  eprint        = {1304.6875},
  archivePrefix = {arXiv},
  primaryClass  = {astro-ph.HE}
}

@article{Cromartie2020,
  author        = {Cromartie, H. T. and Fonseca, E. and Ransom, S. M. and others},
  title         = {Relativistic Shapiro Delay Measurements of an Extremely Massive Millisecond Pulsar},
  journal       = {Nature Astronomy},
  volume        = {4},
  pages         = {72--76},
  year          = {2020},
  doi           = {10.1038/s41550-019-0880-2},
  eprint        = {1904.06759},
  archivePrefix = {arXiv},
  primaryClass  = {astro-ph.HE}
}

@article{Fonseca2021,
  author        = {Fonseca, E. and Cromartie, H. T. and Pennucci, T. T. and others},
  title         = {Refined Mass and Geometric Measurements of the High-Mass {PSR J0740+6620}},
  journal       = {The Astrophysical Journal Letters},
  volume        = {915},
  number        = {1},
  pages         = {L12},
  year          = {2021},
  doi           = {10.3847/2041-8213/ac03b8},
  eprint        = {2104.00880},
  archivePrefix = {arXiv},
  primaryClass  = {astro-ph.HE}
}

@article{Abbott2019GW170817,
  author        = {Abbott, B. P. and others},
  collaboration = {LIGO Scientific Collaboration and Virgo Collaboration},
  title         = {Properties of the Binary Neutron Star Merger {GW170817}},
  journal       = {Physical Review X},
  volume        = {9},
  pages         = {011001},
  year          = {2019},
  doi           = {10.1103/PhysRevX.9.011001},
  eprint        = {1805.11579},
  archivePrefix = {arXiv},
  primaryClass  = {gr-qc}
}

@article{EinsteinTelescope2020,
  author        = {Maggiore, M. and others},
  title         = {Science Case for the {Einstein Telescope}},
  journal       = {Journal of Cosmology and Astroparticle Physics},
  volume        = {2020},
  number        = {03},
  pages         = {050},
  year          = {2020},
  doi           = {10.1088/1475-7516/2020/03/050},
  eprint        = {1912.02622},
  archivePrefix = {arXiv},
  primaryClass  = {astro-ph.CO}
}

@article{CosmicExplorer2019,
  author        = {Reitze, D. and others},
  title         = {{Cosmic Explorer}: The {U.S.} Contribution to Gravitational-Wave Astronomy beyond {LIGO}},
  journal       = {Bulletin of the American Astronomical Society},
  volume        = {51},
  number        = {7},
  pages         = {035},
  year          = {2019},
  eprint        = {1907.04833},
  archivePrefix = {arXiv},
  primaryClass  = {astro-ph.IM}
}

@article{Bauswein2019,
  author        = {Bauswein, A. and Bastian, N.-U. F. and Blaschke, D. B. and Chatziioannou, K. and Clark, J. A. and Fischer, T. and Oertel, M.},
  title         = {Identifying a First-Order Phase Transition in Neutron-Star Mergers through Gravitational Waves},
  journal       = {Physical Review Letters},
  volume        = {122},
  pages         = {061102},
  year          = {2019},
  doi           = {10.1103/PhysRevLett.122.061102},
  eprint        = {1809.01116},
  archivePrefix = {arXiv},
  primaryClass  = {astro-ph.HE}
}

@article{Bernuzzi2020,
  author        = {Bernuzzi, Sebastiano},
  title         = {Neutron Star Merger Remnants},
  journal       = {General Relativity and Gravitation},
  volume        = {52},
  pages         = {108},
  year          = {2020},
  doi           = {10.1007/s10714-020-02752-5},
  eprint        = {2004.06419},
  archivePrefix = {arXiv},
  primaryClass  = {astro-ph.HE}
}

@misc{Zhou2024TwoPeaksSoundSpeedQCD,
  author        = {Zhou, Dake},
  title         = {Bayesian evidence for two peaks in the sound speed in cold dense {QCD}},
  year          = {2024},
  eprint        = {2412.08760},
  archivePrefix = {arXiv},
  primaryClass  = {nucl-th},
  doi           = {10.48550/arXiv.2412.08760},
  url           = {https://arxiv.org/abs/2412.08760}
}

@article{CaiLi2024StrongGravitySoundSpeedPeaks,
  author        = {Cai, Bao-Jun and Li, Bao-An},
  title         = {Strong gravity extruding peaks in speed of sound profiles of massive neutron stars},
  journal       = {Physical Review D},
  volume        = {109},
  number        = {8},
  pages         = {083015},
  year          = {2024},
  doi           = {10.1103/PhysRevD.109.083015},
  eprint        = {2311.13037},
  archivePrefix = {arXiv},
  url           = {https://arxiv.org/abs/2311.13037}
}

@article{YaoEtAl2024StructureSpeedSoundHIC,
  author        = {Yao, Nanxi and Sorensen, Agnieszka and Dexheimer, Veronica and Noronha-Hostler, Jacquelyn},
  title         = {Structure in the speed of sound: From neutron stars to heavy-ion collisions},
  journal       = {Physical Review C},
  volume        = {109},
  number        = {6},
  pages         = {065803},
  year          = {2024},
  doi           = {10.1103/PhysRevC.109.065803},
  eprint        = {2311.18819},
  archivePrefix = {arXiv},
  url           = {https://arxiv.org/abs/2311.18819}
}

@article{MroczekEtAl2024NontrivialFeaturesSoundSpeed,
  author        = {Mroczek, Debora and Miller, M. Coleman and Noronha-Hostler, Jacquelyn and Yunes, Nicolas},
  title         = {Nontrivial features in the speed of sound inside neutron stars},
  journal       = {Physical Review D},
  volume        = {110},
  number        = {12},
  pages         = {123009},
  year          = {2024},
  doi           = {10.1103/PhysRevD.110.123009},
  eprint        = {2309.02345},
  archivePrefix = {arXiv},
  primaryClass  = {astro-ph.HE},
  url           = {https://arxiv.org/abs/2309.02345}
}

@article{EckerRezzolla2022ScaleIndependentSoundSpeed,
  author        = {Ecker, Christian and Rezzolla, Luciano},
  title         = {A General, Scale-independent Description of the Sound Speed in Neutron Stars},
  journal       = {The Astrophysical Journal Letters},
  volume        = {939},
  number        = {2},
  pages         = {L35},
  year          = {2022},
  doi           = {10.3847/2041-8213/ac8674},
  eprint        = {2207.04417},
  archivePrefix = {arXiv},
  primaryClass  = {astro-ph.HE},
  url           = {https://arxiv.org/abs/2207.04417}
}

@article{Douchin2001SLy,
  author  = {Douchin, F. and Haensel, P.},
  title   = {A unified equation of state of dense matter and neutron star structure},
  journal = {Astronomy \& Astrophysics},
  year    = {2001},
  volume  = {380},
  pages   = {151--167},
  doi     = {10.1051/0004-6361:20011402},
  eprint  = {astro-ph/0111092},
  archivePrefix = {arXiv}
}

@article{Typel2015CompOSE,
  author  = {Typel, S. and Oertel, M. and Kl{\"a}hn, T.},
  title   = {CompOSE CompStar online supernova equations of state harmonising the concert of nuclear physics and astrophysics},
  journal = {Physics of Particles and Nuclei},
  year    = {2015},
  volume  = {46},
  pages   = {633--664},
  doi     = {10.1134/S1063779615040061},
  eprint  = {1307.5715},
  archivePrefix = {arXiv},
  primaryClass = {astro-ph.SR}
}

@article{Sorkin1981,
  author  = {Sorkin, Rafael D.},
  title   = {A Criterion for the Onset of Instability at a Turning Point},
  journal = {The Astrophysical Journal},
  year    = {1981},
  volume  = {249},
  pages   = {254--257},
  doi     = {10.1086/159282}
}

@article{Friedman1988,
  author  = {Friedman, John L. and Ipser, James R. and Sorkin, Rafael D.},
  title   = {Turning-Point Method for Axisymmetric Stability of Rotating Relativistic Stars},
  journal = {The Astrophysical Journal},
  year    = {1988},
  volume  = {325},
  pages   = {722--724},
  doi     = {10.1086/166043}
}

@article{Chandrasekhar1964,
  author  = {Chandrasekhar, Subrahmanyan},
  title   = {Dynamical Instability of Gaseous Masses Approaching the Schwarzschild Limit in General Relativity},
  journal = {The Astrophysical Journal},
  year    = {1964},
  volume  = {140},
  pages   = {417--433},
  doi     = {10.1086/147938}
}

@article{Bardeen1966,
  author  = {Bardeen, James M. and Thorne, Kip S. and Meltzer, David W.},
  title   = {A Catalogue of Methods for Studying the Normal Modes of Radial Pulsation of General-Relativistic Stellar Models},
  journal = {The Astrophysical Journal},
  year    = {1966},
  volume  = {145},
  pages   = {505--513},
  doi     = {10.1086/148791}
}

@article{Kokkotas2001,
  author        = {Kokkotas, Kostas D. and Ruoff, Johannes},
  title         = {Radial Oscillations of Relativistic Stars},
  journal       = {Astronomy \& Astrophysics},
  year          = {2001},
  volume        = {366},
  pages         = {565--572},
  doi           = {10.1051/0004-6361:20000216},
  eprint        = {gr-qc/0011093},
  archivePrefix = {arXiv}
}

@misc{Moffat2020MOGGW190814,
  title         = {Modified Gravity ({MOG}) and Heavy Neutron Star in Mass Gap},
  author        = {Moffat, J. W.},
  year          = {2020},
  archivePrefix = {arXiv},
  eprint        = {2008.04404},
  primaryClass  = {gr-qc},
  doi           = {10.48550/arXiv.2008.04404}
}

@article{Astashenok2020ExtendedGravityGW190814,
  title         = {Extended gravity description for the {GW190814} supermassive neutron star},
  author        = {Astashenok, Artyom V. and Capozziello, Salvatore and Odintsov, Sergei D. and Oikonomou, Vasilis K.},
  journal       = {Physics Letters B},
  volume        = {811},
  pages         = {135910},
  year          = {2020},
  doi           = {10.1016/j.physletb.2020.135910},
  archivePrefix = {arXiv},
  eprint        = {2008.10884},
  primaryClass  = {gr-qc}
}

@article{Nunes2020ScreeningGravityGW190814,
  title         = {Weighing massive neutron star with screening gravity: a look on {PSR J0740+6620} and {GW190814} secondary component},
  author        = {Nunes, Rafael C. and Coelho, Jaziel G. and {de Araujo}, Jose C. N.},
  journal       = {European Physical Journal C},
  volume        = {80},
  pages         = {1115},
  year          = {2020},
  doi           = {10.1140/epjc/s10052-020-08695-0},
  archivePrefix = {arXiv},
  eprint        = {2008.10395},
  primaryClass  = {astro-ph.HE}
}

@article{Lin2022FTGravityGW190814,
  title         = {Realistic neutron star models in {$f(T)$} gravity},
  author        = {Lin, Rui-Hui and Chen, Xiao-Ning and Zhai, Xiang-Hua},
  journal       = {European Physical Journal C},
  volume        = {82},
  pages         = {308},
  year          = {2022},
  doi           = {10.1140/epjc/s10052-022-10268-2},
  archivePrefix = {arXiv},
  eprint        = {2109.00191},
  primaryClass  = {gr-qc}
}

@article{Charmousis2022EGBGW190814,
  title         = {Astrophysical constraints on compact objects in {4D} {Einstein--Gauss--Bonnet} gravity},
  author        = {Charmousis, Christos and Leh{\'e}bel, Antoine and Smyrniotis, Evangelos and Stergioulas, Nikolaos},
  journal       = {Journal of Cosmology and Astroparticle Physics},
  volume        = {2022},
  number        = {02},
  pages         = {033},
  year          = {2022},
  doi           = {10.1088/1475-7516/2022/02/033},
  archivePrefix = {arXiv},
  eprint        = {2109.01149},
  primaryClass  = {gr-qc}
}

@article{Reyes2024PostTOVScreenedGW190814,
  title         = {Parameterized Post-{Tolman--Oppenheimer--Volkoff} Framework for Screened Modified Gravity with an Application to the Secondary Component of {GW190814}},
  author        = {Reyes, Christopher and Sakstein, Jeremy},
  journal       = {Physical Review D},
  volume        = {109},
  pages         = {084080},
  year          = {2024},
  doi           = {10.1103/PhysRevD.109.084080},
  archivePrefix = {arXiv},
  eprint        = {2403.03399},
  primaryClass  = {gr-qc}
}

\end{document}